\documentclass[fleqn,usenatbib]{mnras}

\usepackage{newtxtext,newtxmath}

\usepackage{lineno}

\usepackage[T1]{fontenc}

\DeclareRobustCommand{\VAN}[3]{#2}
\let\VANthebibliography\thebibliography
\def\thebibliography{\DeclareRobustCommand{\VAN}[3]{##3}\VANthebibliography}

\usepackage{graphicx}	
\usepackage{amsmath}	

\usepackage{soul}
\usepackage[dvipsnames]{xcolor}

\title[Bias from Correlated Observable Selection]{The effect of selection -- a tale of cluster mass measurement bias induced by correlation and projection}

\author[]{
Yuanyuan Zhang$^{1, 2}$\thanks{E-mail: ynzhang@tamu.edu}, James Annis$^{2}$
\\
$^{1}$ Mitchell Institute for Fundamental Physics and Astronomy and Department of Physics and Astronomy, Texas A\&M University, College Station, TX 77843-4242\\
$^{2}$ Fermilab, Cosmic Physics Center, Kirk\& Pine Road, Batavia, IL, 60510 \\
}
\date{Accepted XXX. Received YYY; in original form ZZZ}

\pubyear{2015}

\begin{document}
\label{firstpage}
\pagerange{\pageref{firstpage}--\pageref{lastpage}}
\maketitle

\begin{abstract}
Cosmology analyses using galaxy clusters by
the Dark Energy Survey  have recently uncovered an issue of previously unknown selection effect  affecting weak lensing mass estimates.
In this letter, we use the Illustris-TNG simulation to demonstrate that selecting on galaxy counts induces a selection effect because of projection and correlation between different observables. We compute the weak-lensing-like projected mass estimations of dark matter halos, and  examine their projected subhalo counts. In the 2-D projected space, halos that are measured as more massive than  truth have higher subhalo counts. Thus, projection along the line of sight creates cluster observables that are correlated with cluster mass measurement deviations, which in turn creates a mass measurement bias when the clusters are selected by this correlated observable. We  demonstrate that the bias is predicted in a forward model using the  observable-mass measurement correlation. 
\end{abstract}




\section{Introduction}

Galaxy clusters have long been utilized as sensitive probes of cosmology in optical wide-field surveys such as the Sloan Digital Sky Survey \citep{2010ApJ...708..645R,2014MNRAS.439.1628Z, 2019MNRAS.488.4779C}, the Dark Energy Survey \citep{2020PhRvD.102b3509A} and also planned for the upcoming Legacy Survey of Space and time at the Rubin Observatory \citep{2018arXiv180901669T}. These optical survey programs study galaxy clusters by their optical observables \citep[e.g.,][]{2014ApJ...785..104R, 2020MNRAS.493.4591P} and weak lensing mass signals \citep{2017MNRAS.466.3103S, 2019MNRAS.482.1352M, 2020MNRAS.498.5450P}, often to a lower mass threshold than X-ray and CMB experiments \citep[e.g.,][]{2009ApJ...692.1060V, 2015MNRAS.446.2205M, 2016A&A...594A..24P, 2019ApJ...878...55B}. However recent discoveries from the Dark Energy Survey in \cite{2020PhRvD.102b3509A} point to additional systematic effects that plague the accuracy of cluster cosmological constraints, which stem from previously undetected selection effects biasing cluster weak lensing mass measurements. Ongoing and future cosmic surveys will need to address this challenge in order to achieve the full potential of galaxy cluster cosmology analyses. 

Specifically, the analysis in \cite{2020PhRvD.102b3509A} shows that the weak-lensing measured masses of galaxy clusters appear to have deviated from their model values significantly, which, in the end, affect cosmological parameters derived from modeling the cluster mass distribution and abundance. 
\cite{2020PhRvD.102b3509A} and further \cite{2020MNRAS.496.4468S} find that the bias of cluster weak-lensing measurements likely originate from cluster selection. Their analyses based on simulations have demonstrated that galaxy clusters selected by their richness observable, defined as a weighted number count of red sequence cluster galaxies, is a biased population when compared to a unbiased population selected by their unbiased truth quantities. \cite{2020MNRAS.496.4468S} has further determined that the projection of cosmic structures along the line of sight, as well as cluster orientation and shapes \citep[][and Z. Zhang in prep.]{2014MNRAS.443.1713D,2018MNRAS.477.2141O} may be causing this selection bias.

These effects may cause additional correlation between the cluster's observable and its weak-lensing mass measurement, and thus we can consider alleviating their effects through modeling the correlations \citep[ as in, e.g., ][]{2019MNRAS.487.2578Z, 2020MNRAS.498..771G, 2021MNRAS.504.1253G}.
In this paper, we demonstrate the potential effectiveness of this correlation approach by examining the projected observables of cluster-sized dark matter halos and their correlations with weak-lensing like mass measurements in simulation. Through doing so, we validate previous conclusions about the "projection" origin of the previously over-looked selection effect.

\section{Simulation Data}
\label{sec:sims} 

\begin{figure}
	\includegraphics[width=0.9\columnwidth]{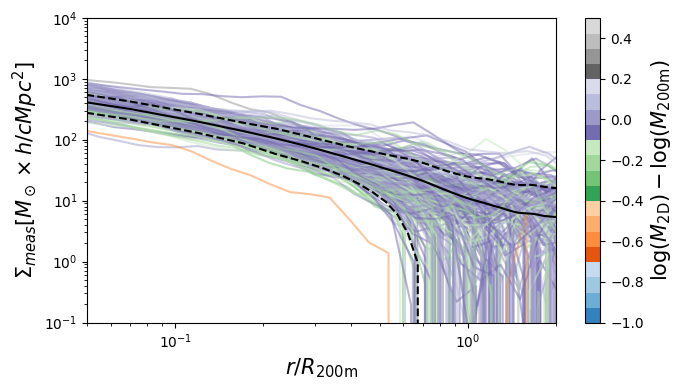}
    \caption{Projected mass surface density profiles of the massive dark matter halos studied in this paper. The color coding indicate their weak-lensing-like mass measurement deviations shown in Figure 2. The solid and dashed black lines show the mean and the 1 $\sigma$ uncertainty range of the distributions.}
    \label{fig:covariance}
\end{figure}

In this work, we use products from the IllustrisTNG simulation suite, in particular, from the IllustrisTNG 300-1 simulation \citep{2018MNRAS.475..648P, 2018MNRAS.475..624N, 2018MNRAS.475..676S, 2018MNRAS.477.1206N, 2018MNRAS.480.5113M, 2019ComAC...6....2N}, the largest hydro-dynamic simulation box, with high-resolution baryonic processes incorporated. We analyze dark matter halos in the $79th$ snapshot, corresponding to redshift 0.27, which is the intermediate redshift of the lowest redshift bin used in DES cluster cosmology analysis. 
We have also analyzed the simulation snapshots that correspond to redshift 0.42 and 0.58, and find that similar effects reported in this letter are also present.

We analyze halos with $M_\mathrm{200m}$ above $10^{13.75} M_\odot/h$ \footnote{Our analysis of lower mass halos, especially those below $10^{13.5} M_\odot/h$, shows that the adopted halo models in this paper do not work well, which may be worthy of a separate study itself.}. For each of the halo, we derive their density profiles using the dark matter, gaseous and stellar particles in the simulation snapshot. These particles are projected onto the $X-Y$ plane of the simulation with a projection depth along the $Z$ direction of 120 $\mathrm{cMpc}/h$ \footnote{The letter $c$ indicates comoving distance as a notation adopted in IllustrisTNG simulation.}.
Given that the the simulation box size is 205 $\mathrm{cMpc}/h$ on each side, to maximize the range of projection depth,
we only use the particles that are located on one side of the halo's $Z$-axis up to 120 $\mathrm{cMpc}/h$ from the halo center. The projected density profile is derived according to the dark matter, stellar and gaseous particles' projected distance to the halo center on the $X-Y$ plane, and then multiplied by a factor of 2 assuming the halo to be symmetrical on the $Z$-axis to recover its total density profile. A projected background density is estimated as the averaged density of the dark matter, stellar and gaseous particles projected onto the $X-Y$ plane, with a thickness of 240 cMpc/h along the $Z$-axis. 
To avoid running into the simulation boundaries, we exclude halos that are within 20 cMpc/h of the simulation box, or 20 $\mathrm{cMpc/h}$ within the the $Z$-axis middle plane ($z=102.50 \mathrm{~cMpc/h}$). In the end, we analyze a total of 254 dark matter halos. 

We have investigated the derived halo matter profiles are sensitivities to (1) baryonic effects (2) simulation resolution and (3) projection depth. 
In the IllustrisTNG dark-matter only simulation, TNG-Dark-300-1, the density profiles of halos above $10^{14} M_\odot/h$ show a relative difference of up to $\sim 10 \%$ in the central 200 $\mathrm{cKpc}/h$ region when compared to halo profiles from the TNG 300-1 hydrodynamic simulation.
Similarly, when we compare the halo density profiles of our fiducial measurements from the TNG300-1 hydrodynamic simulation to profiles from the lower resolution TNG 300-3, TNG 300-2 simulations, we find differences up to $\sim 5 \%$ within the central 100 $\mathrm{cKpc}/h$ regions. 
Given the accuracies of previous cluster mass calibration studies \citep{2011ApJ...740...25B, 2021arXiv210316212G}, we would like profiles insensitive to baryonic effects or simulation resolution at the 10\% or ideally 5\% level.
Thus we exclude the central 200 $\mathrm{cKpc}/h$ regions of the halo profiles during the further analysis. 

Our investigation of projection depth sets the outer radius. We compute the projected halo density profiles and corresponding projected background densities for projection length, 75, 100 and 135 $\mathrm{cMpc}/h$ from the TNG300-1 and TNG300-3 simulations.  In the central 2 $\mathrm{cMpc}/h$ regions, the halos' profiles agree within $\sim 10\%$ for projection depth $> 100\mathrm{cMpc}/h$.
Outside the 2 $\mathrm{cMpc}/h$ radial range, using different projection depths causes noticeable fluctuations, and the amplitude of fluctuations decreases as  halo masses increase. It is unclear if the halo density profiles agree to within $10\%$ relative difference.
Therefore, in this work, we only analyze the 0.2 to 2 $\mathrm{cMpc}/h$ radial range of the halo density profiles.

The derived halo density profiles are shown in Fig.~\ref{fig:covariance}. We also derive the covariance matrix  as the halo-to-halo variation after normalizing the halo profiles by their radii, the $R_{200m}$ value of each halo \citep[see studies in][]{2019MNRAS.490.2606W}. This covariance matrix is later used during the derivation of halo masses from their projected density profiles. 

\section{Halo Mass Observables}
~\label{sec:mass}

\begin{figure}
	\includegraphics[width=1.0\columnwidth]{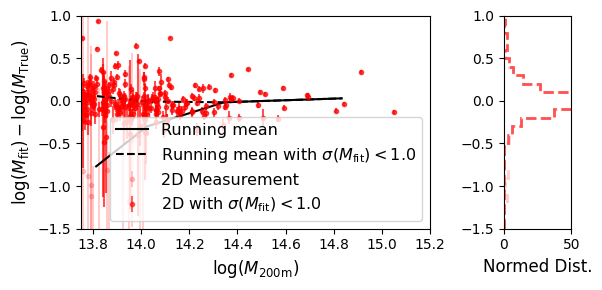}
    \caption{The recovered halo masses, and their 1 $\sigma$ uncertainties (standard deviation from the MCMC posterior sampling) by fitting the projected halo mass density profiles with analytical models.  Note that there is a fraction of dark matter halos below $M_{200m}$ of $10^{14} \mathrm{M_\odot}/h$ that we are unable to reliably recover their masses as the fitting values of concentration and mass has failed to converge, indicated by the data points with large error bars (light-colored data points). For the rest of the halos (dark-colored data points), the analytical models recover the true masses with an average of 1.4\% bias.}
    \label{fig:masses}
\end{figure}


We first derive halo masses in the projected $X-Y$plane. 
We fit the projected halo profiles to a model corresponding to both a halo's gravitationally-bound matter distribution and the contribution from large scale structures, using the following form:
\small
\begin{equation}
\begin{split}
\Sigma(r|M_{200m}, c, z) & = \mathrm{max}(\Sigma_\mathrm{NFW}(r|M_{200m}, c), \Sigma_\mathrm{2-halo}(r|M_{200m}, z)), \\ 
\Sigma_\mathrm{2-halo}(r|M_{200m}, z) & = b(M_{200m}, z) \Sigma_\mathrm{nl}(r|M_{200m}, z) \\
\Sigma_\mathrm{nl}(r|M_{200m}, z) & = \int^{+\infty}_{-\infty} \mathrm{d} \chi \rho_m \xi_\mathrm{nl}(\sqrt{r^2+\chi^2}).
\end{split}
\label{eq:mass_meas}
\end{equation}
\normalsize
Here we combine the projected NFW halo profiles, the so-called "one-halo" term, and the nonlinear matter correlation function weighted by halo bias $b(M_{200m}, z)$, the "two-halo" term. In this equation, $\xi_\mathrm{nl}$ is the non-linear matter correlation function, which is  the 3D Fourier transformation of the non-linear power spectrum. 
These fitting models are inspired by those adopted in \citet{2017MNRAS.466.3103S, 2019MNRAS.482.1352M} and tested to give similar mass estimations within $\sim1\%$ but with improved speed. During the fitting procedure,  the model masses and concentrations are treated as varying parameters sampled through Markov Chain Monte Carlo (MCMC), which is implemented using EMCEE \citep{2013PASP..125..306F} with a likelihood constructed from the $\chi^2$ value between the model and each halos's measurements. The constrained values of the halo's masses are shown in Fig.~\ref{fig:masses} and compared to each halo's truth values.

There are some dark matter halos below $\mathrm{log}M_{200m}$ of $10^{14} \mathrm{M_\odot}/h$ for which we are unable to recover their masses as the fittings of concentration and mass fail to converge to reasonable values. In Fig.~\ref{fig:masses}, those halos have large mass uncertainties, exceeding 1.0 dex.  We were, however, able to recover the masses of those halos using their 3D density profiles, and those halos also tend to have a projected density profile significantly below the average, as shown in Figure \ref{fig:covariance}. This indicates that they may have under-dense environments and thus their profiles are not represented by the analytical models in the 3D to 2D projection process. For the rest of the halos, the analytical models recover the true masses with an average bias of 1.4\% dex.


\begin{figure*}
	\includegraphics[width=0.99\columnwidth]{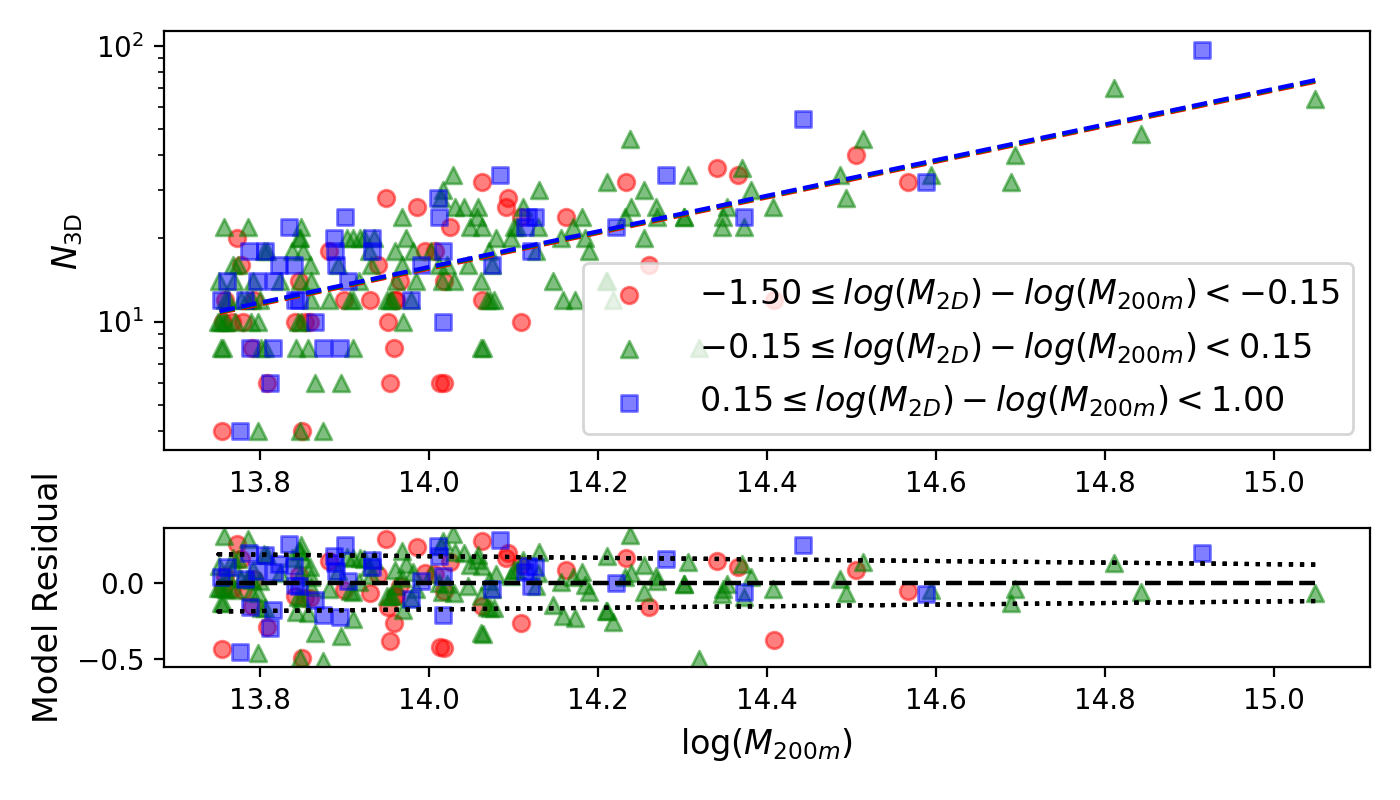}
	\includegraphics[width=0.99\columnwidth]{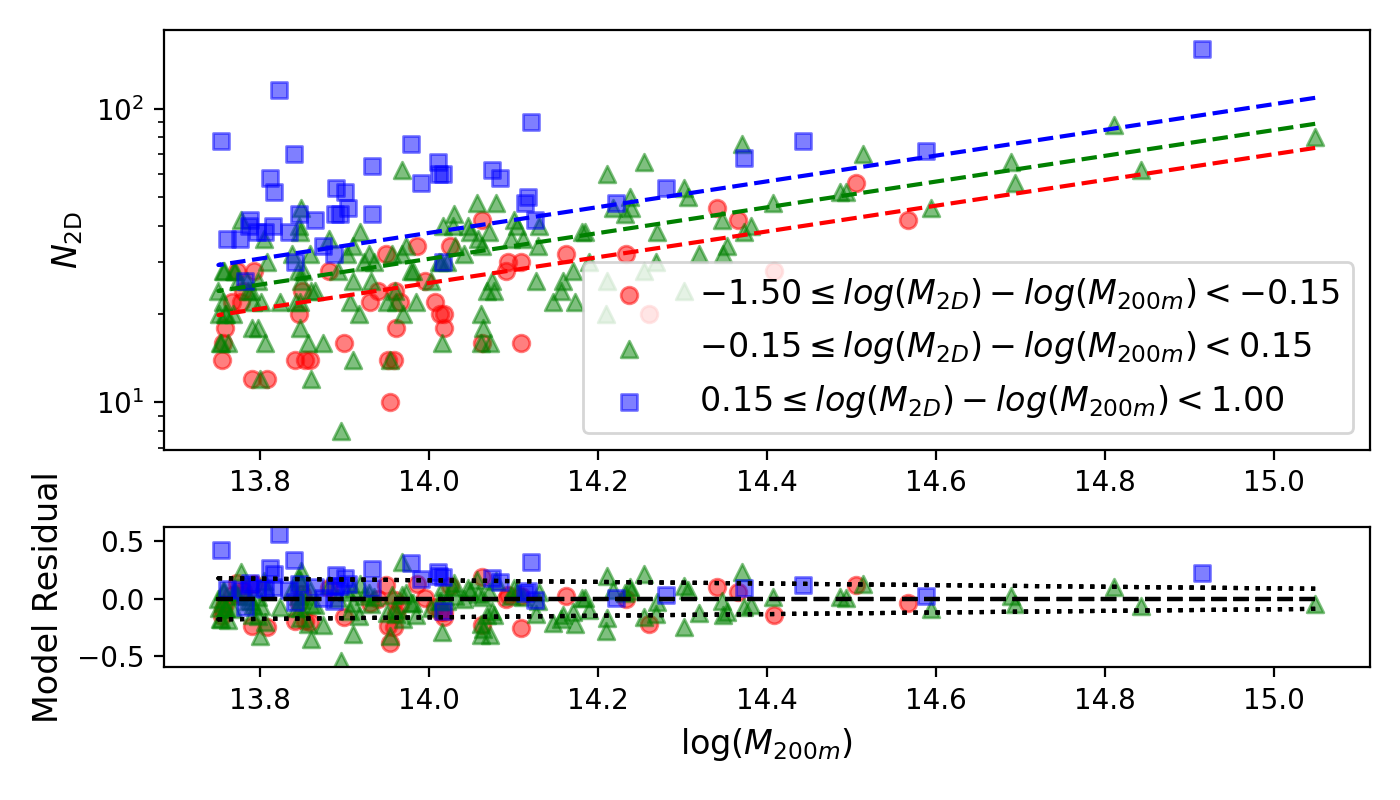} 
    \caption{Relations between subhalo counts, $N_\mathrm{3D}$(left), $N_\mathrm{2D}$ (right)  and halo's masses, color coded by the halo's mass measurement deviations (red, green and blue). $N_\mathrm{2d}$ displays signs of correlation with the halo's mass deviations -- halos that  have positive mass deviations (blue squares) tend to have higher subhalo counts than the average, and vice versa for the halos that have negative mass deviations (red rounds). $N_\mathrm{3D}$ displays no sign of the correlation. The dashed lines in the upper panels indicate the fitted model predictions for the halos with mass deviations in different value ranges, and the lower panels show the residuals between the measured subhalo counts and their mean model values, with the dotted lines showing the modelled scatters.  }
    \label{fig:corr}
\end{figure*}

In this section, we compute the number of subhalos in dark matter halos as a halo mass proxy. 
In optical studies, the cluster selection process often relies on a galaxy over-density above a given threshold.
In this exercise, we aim to reproduce this selection by counting the number of massive subhalos with masses above $5\times10^{9} \mathrm{M_\odot}/h$, within a radial aperture around the halo centers. We consider two different kinds of apertures based on the 1 $\mathrm{Mpc}$ physical distance radius (close to the average $R_\mathrm{200m}$ analyzed of the halo samples, which is 0.87 cMpc/h): 
\begin{itemize}
    \item A 3D-radial aperture of 1 $\mathrm{cMpc}/h$. With this aperture, any massive subhalos that have a 3-D distance less than 1 $\mathrm{cMpc}/h$ to the halo center are counted. The number of subhalos satisfying this criteria is designated $N_\mathrm{3D}$.
    \item A 2D-radial aperture of 1 $\mathrm{cMpc}/h$. With this aperture, massive subhalos that have a 2-D distance (in the $X-Y$ plane of the simulation) less  than 1 $\mathrm{cMpc}/h$ to the halo center are counted. Only subhalos within a given projection length selected in the same way of the mass profile projection in Sect.~\ref{sec:sims} are considered. The number of subhalos is designated $N_\mathrm{2D}$.
\end{itemize}

In Fig.~\ref{fig:corr}, we demonstrate that the halo's mass measurement deviations are correlated with subhalo number counts, when controlling the halo's truth mass. In this figure, the subhalo number counts of $N_\mathrm{3D}$ and $N_\mathrm{2D}$ are shown against the halo's truth mass, showing a clear relation between the two: more massive halos host more subhalos. More interestingly, the relation between the subhalo counts and halo mass is further correlated with the halo's mass measurement deviations (derived in the previous sub-section in the case of $N_\mathrm{2D}$). Halos that are measured as more massive than their truth masses  have a higher $N_\mathrm{2D}$ values (blue data points), while halos that are measured to be less massive than their truth masses have lower $N_\mathrm{2D}$ values (red data points).

To further quantify this correlation, we paramterize the relation between subhalo counts and the halo's masses using the following statistical model for counts as a function of mass and mass deviation: 
\small
\begin{equation}
\label{eq:om_relation}
\begin{split}
   \mathrm{log} N &= N(\mu(M_\mathrm{true}, M_\mathrm{fit}), \sigma^2(M_\mathrm{true}, M_\mathrm{fit})) \\
 \Delta & = \mathrm{log}M_\mathrm{fit} - \mathrm{log}M_\mathrm{true} \\
   \mu(M_\mathrm{true}, M_\mathrm{fit}) & = a\times(\mathrm{log}M_\mathrm{true}-14.0) + b + (\alpha + \beta \times \Delta )\times\Delta  \\
   \sigma(M_\mathrm{true}, M_\mathrm{fit}) & = \sigma_0 + q \times (\mathrm{log}M_\mathrm{true} - 15.5)
\end{split}
\end{equation}
\normalsize
This model has a linear relation between the subhalo counts and halo masses at the log scale as adopted in some cluster cosmology studies \citep[e.g.,][]{2010ApJ...708..645R,2014MNRAS.439.1628Z}, although newer studies suggest a more complicated functional forms to be more accurate \citep{2019MNRAS.482..490C, 2019MNRAS.488.4779C}. The subhalo counts  deviations from the mean are described by a Gaussian scatter, and the Gaussian scatter depends on the halo mass in a linear form as adopted in \cite{2018ApJ...854..120M}. However in order to quantify the additional correlation between $N_\mathrm{2D}$ and the halo mass measurement deviation, we further incorporate first-order and second-order dependencies on the mass measure deviations $\Delta  = \mathrm{log}M_\mathrm{fit} - \mathrm{log}M_\mathrm{true}$,  in the mean linear relation, controlled by the values of parameters $\alpha$ and $\beta$.  We have attempted incorporating these dependencies in the $N_\mathrm{2D}$ Gaussian scatters, but do not find significant dependencies and therefore do not adopt them here. The free parameters in the relations, $a$, $b$, $\alpha$, $\beta$, $\sigma_0$ and $q$ are constrained using Markov Chain Monte Carlo sampling with flat Bayesian priors. 

\begin{table*}
	\centering
	\caption{Constraints on the parameters of Eq.~\ref{eq:om_relation}, the halo's subhalo counts observable to mass relation, for different subhalo count definitions.}
	\label{tab:table}
	\begin{tabular}{lccccccr} 
		\hline
	Model &	$a$ & $b$ & $\alpha$ & $\beta$ & $\sigma_0$ & q  & $\chi^2$ ($n=254$) \\
		\hline
		$N_\mathrm{3D}$ & $0.65 \pm 0.04$ & $1.195 \pm 0.011$ & $0.009\pm 0.033$ & $-0.003 \pm 0.006$ & $0.096 \pm 0.038$    & $-0.052 \pm 0.026$ & 245.38\\
		$N_\mathrm{2D}$ & $0.439 \pm 0.037$ & $1.491 \pm 0.011$ & $0.259 \pm 0.031$ & $0.036 \pm 0.006$ & $0.056 \pm 0.032$   & $-0.069 \pm 0.022$ & 245.35 \\
		$N_\mathrm{r3D}$ &  $0.916 \pm 0.035$ & $1.23 \pm 0.011$ & $0.011 \pm 0.032$ & $-0.002 \pm 0.006$  & $0.043 \pm 0.033$   & $-0.086 \pm 0.022$ & 245.71\\
		$N_\mathrm{r2D}$ &  $0.750 \pm 0.035$ & $1.56 \pm 0.011$ & $0.268 \pm 0.031$ & $0.038 \pm 0.006$  & $0.041 \pm 0.030$   & $-0.070 \pm 0.020$ & 245.42\\
		$N_\mathrm{iterR}$ &  $0.735 \pm 0.056$ & $1.53 \pm 0.018$ & $0.414 \pm 0.053$ & $0.062 \pm 0.010$  & $0.056 \pm 0.044$   & $-0.146 \pm 0.030$ & 246.43\\
		\hline
	\end{tabular}
\end{table*}

The posterior values of the constrained parameters are listed in Table~\ref{tab:table}. Interestingly, $N_\mathrm{2D}$ have significant correlation with the mass measurement deviation, given the positive values of $\alpha$ and $\beta$. On the other hand, $N_\mathrm{3D}$  does not have a significant correlation with the mass measurement deviation and the values of $\alpha$ and $\beta$ are consistent with 0. This result is in agreement with the qualitative observations in Fig.~\ref{fig:corr}.

We further investigate this effect. In observational studies, the radial apertures used to derive a cluster galaxy over-density observable is often iteratively adjusted according to the cluster's ``sizes''/``masses''. Thus we further test the correlations with the following subhalo counting definitions that use apertures based on halo masses:
\begin{itemize}
    \item A 3D-radial aperture of $R_\mathrm{200m}$, derived from the halo's $M_\mathrm{200m}$. Any massive subhalos that have a 3-D distance less than $R_\mathrm{200m}$ to the halo centers are counted. The number of subhalos satisfying this criteria is designated $N_\mathrm{r3D}$.
    \item A 2D-radial aperture of $R_\mathrm{200m}$, derived from the halo's $M_\mathrm{200m}$. Any massive subhalos that have a 2-D distance (in the X-Y plane of the simulation) less than $R_\mathrm{200m}$ to the halo centers are counted. The number of subhalos satisfying this criteria is designated $N_\mathrm{r2D}$.
    \item An iterative 2D-radial aperture $R_\mathrm{iter}$, which is adjusted according to the derived subhalo counts $N_\mathrm{iterR}$ until they satisfy the relation $N_\mathrm{iterR} = 26.514\times(R_\mathrm{iter}/\mathrm{[cMpc/h]})^{2.414} $. This relation is the mean relation between  $R_\mathrm{200m}$ and $N_\mathrm{r2D}$ defined above.
\end{itemize} 
The last is designed to model the usual case in optical observations that a cluster's true $R_\mathrm{200m}$ is not known, but the galaxy counting apertures can be iteratively improved from the counts.

The parameterization of the relations between those subhalo counts, and halo masses are also listed in Table~\ref{tab:table}. Again, the 3D-radial aperture based observable, $N_\mathrm{r3D}$, shows no correlation with the halo mass measurement deviation. On the other hand, $N_\mathrm{iterR}$ has the strongest correlation with the mass measurement deviation. 
We expect being in a high-density environment, or other similar factors, affect both halo mass measurements and the apertures used to select galaxy over-densities. This creates the greater correlation between the two quantities as seen with $N_\mathrm{r3D}$.


\section{Mass Bias of the Selected Halos}

When there exist correlations between the halos's observed mass and their selection observable, the selected halo sample exhibits a  mass measurement bias. Assuming that the halo's observed mass is a fractional deviation from the halo's truth mass, the mathematical expression of the bias can be derived by examining the average masses of the halos selected by observable,

\scriptsize
\begin{equation}
    \begin{split}
       M_{M}(N) &=   \frac{1}{P(N)}\int_{-\infty}^{\infty} \mathrm{d} \Delta \int_{10^{13.75}}^{\infty} \mathrm{d}M (M10^{\Delta}) P(N|M, \Delta) P(\Delta | M) P(M)  \\
       M_{T}(N) &= \frac{1}{P(N)} \int_{-\infty}^{\infty} \mathrm{d} \Delta \int_{10^{13.75}}^{\infty} \mathrm{d}M (M) P(N|M, \Delta)  P(\Delta | M) P(M) \\
       \text{Bias} & = \mathrm{log}\frac{M_{M}(N)}{M_{T}(N)}\\
    \end{split}
\end{equation}
\normalsize
In those equations, $M_M$ represents the average measured mass of the halos selected by the observable $N$, while $M_T$ represents their truth average mass. In computing those biases, $P(N|M, \Delta)$ is the halo's mass-observable relation, incorporating $N$'s correlation to halo masses and measurement deviations studied in Section 3, while $P(M)$ is the halo's true mass distribution, derived from the theoretical halo mass function. $P(\Delta | M)$ models the distribution of individual halo mass measurement deviations without selection, which has been studied previously in literature in the context of "mass calibration" for cluster weak lensing studies \citep{2011ApJ...740...25B}. Here, we model it as a Log-Normal distribution, dependent on the halo's truth mass as described in Sect. ~\ref{sec:mass}. We impose a requirement that $ \int_{-\infty}^{\infty} \mathrm{d} \Delta \mathrm{10}^{\Delta} P(\Delta|M) = 1$, by adjusting the mean value of $\Delta$ based on the measured scatter of $\Delta$ so that when there is no correlation between $N$ and $\Delta$, $P(N|M, \Delta)$ reduces to $P(N|M)$, and $M_M(N) =  M_N(N)$. In the situation that $N$ correlates to $\Delta$ in $P(N|M, \Delta)$, then  $M_M(N)$ and $M_N(N)$ are no longer necessarily equivalent.

\begin{figure}
	\includegraphics[width=1.1\columnwidth]{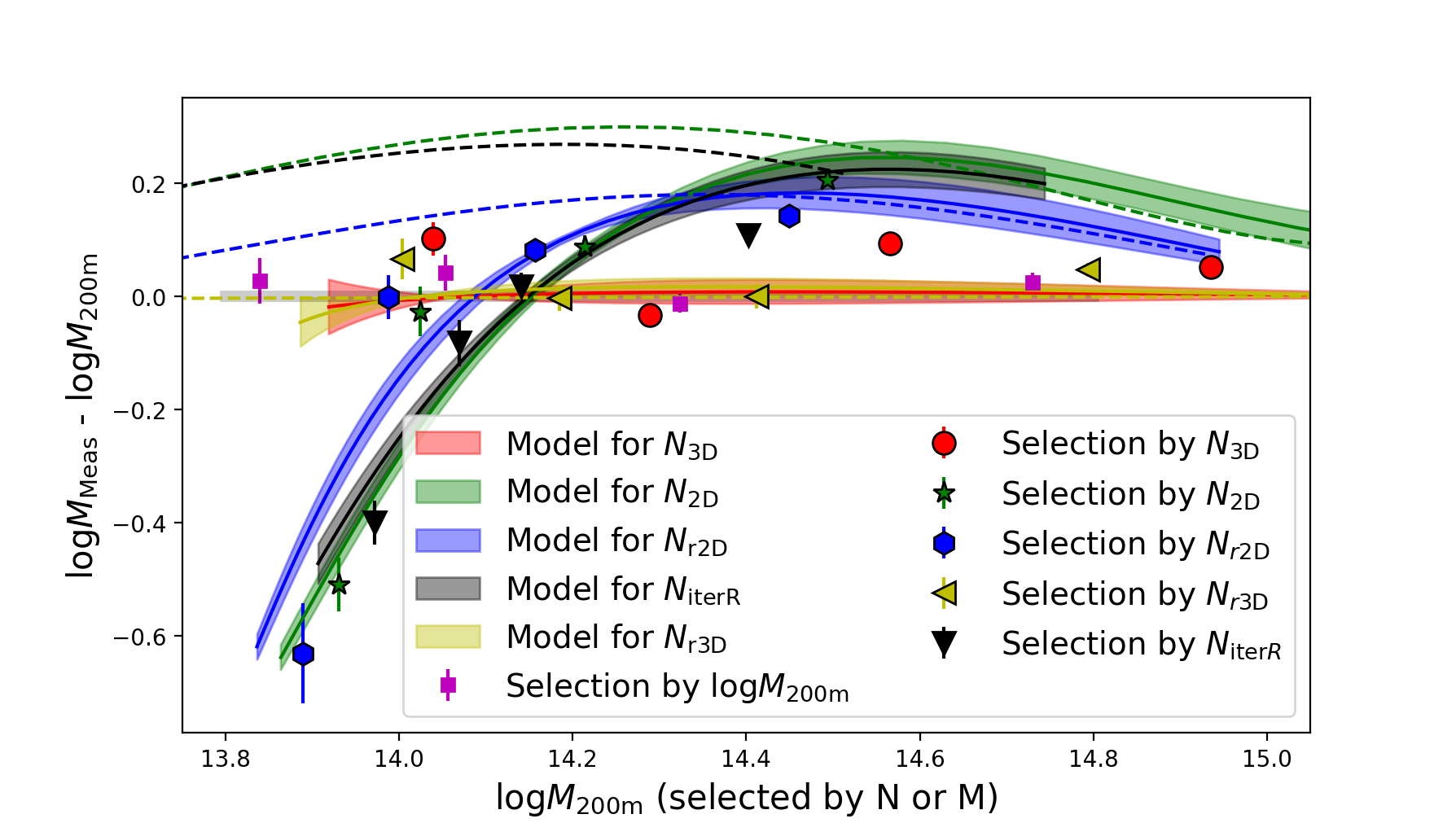}
    \caption{Bias in the measurement of the clusters' average masses while the clusters are selected by different mass observables. Clusters selected by the projected subhalo counts ($N_\mathrm{2D}$, dash-dotted green line), which has strong observable-mass deviation correlation, show a high level of measurement bias, while the clusters selected by 3D subhalo counts ($N_\mathrm{3D}$, dashed red line) or true masses ($\mathrm{log}M_\mathrm{200m}$, solid black line) show no significant biases.} 
    \label{fig:bias}
\end{figure}

We quantify the mass measurement bias of halos selected by $N$ as $\mathrm{log}\frac{M_{M}(N)}{M_{T}(N)}$.
Using quantities derived in previous sections and a Tinker mass function \citep{2008ApJ...688..709T} 
as $P(M)$, we show the bias prediction $\mathrm{log}\frac{M_{M}}{M_{T}}$ in Fig.~\ref{fig:bias}. When the selection observable $N$ has no significant correlation with the halo mass measurement deviation, as is the case with $N_\mathrm{3D}$ and $N_\mathrm{r3D}$, the derived biases are consistent with 0 (shaded red and yellow bands respectively). When the selection observable are correlated, as are the cases with $N_\mathrm{2D}$, $N_\mathrm{r2D}$, and $N_\mathrm{iterR}$, the theoretically derived biases (shaded green, blue and magenta bands) are no longer consistent with 0. Note that because of the artificially-imposed mass selection cut at $10^{13.75}\mathrm{M_\odot}/h$, the mass measurement bias turns negative at the lower mass end in both measurement and theoretical calculation because of the lack of up-scattered halos from lower mass ranges. This trend would disappear if we removed the halo mass selection threshold (dashed lines). In general, the more correlated observable yield more prominent mass biases when selected upon, and the biases also vary with halo masses.

Do those expectation match what we would observe in observations? 
We further measure the mass measurement biases of the halos selected according to the different subhalo counts.  We group the halos according to their subhalo counts and measure their average masses, a technique often employed by cluster lensing analyses. The average masses are measured by fitting Equations~\ref{eq:mass_meas} to their averaged radial profiles with a single set of mass and concentration values. This method does assume that the averaged halo profiles are well described by the model of a single halo, but this has been tested to be accurate at the 5\% level \citep{2017MNRAS.469.4899M, 2019MNRAS.482.1352M} for recovered halo masses.

Fig.~\ref{fig:bias} shows the results of the recovered average halo mass selected by the different observables, in terms of their measurement biases {\it vs} their truth average mass. The halos selected by $N_\mathrm{2D}$, $N_\mathrm{r2D}$, and $N_\mathrm{iterR}$ show biases in their averaged mass measurements. This is in contrast to the halos that are grouped according to their truth masses $\mathrm{log}M_\mathrm{200m}$ and $N_{3D}$, $N_\mathrm{r3D}$, which show a negligible level of biases. Those biases qualitatively match the quantitative expectations from the parameterized models, and in the high mass end, also match the effect of selection biases discovered in \cite{2020PhRvD.102b3509A, 2020MNRAS.496.4468S} at the $\sim 10\%$ level. 

In observational applications, the exact forms of those correlations will not need to be precisely known. In cluster cosmological analysis, the cluster mass-observable relations are modeled as nuisance parameters together with cosmological parameters, which can be expanded to include correlations and simultaneously constrained with the cluster cosmological parameters. Our analysis indicate that including the correlation parameters has the potential to minimize the selection bias plaguing the precision of cluster cosmology analysis as discussed in \cite{2020PhRvD.102b3509A}. An alternative for tackling the selection bias is to rely on cluster observables that are either not or are less affected by projection effect to select cluster samples, as demonstrated by the mass bias results based on $N_{3D}$, $N_\mathrm{r3D}$. 

\section{Summary}
In this letter, we use the TNG simulation to explore the effects of cluster selection on  weak-lensing-like mass measurements. We demonstrate that 
when dark matter halo are selected by their projected galaxy-like observables, these observables become correlated with the halo weak-lensing-like mass measurement deviations because weak lensing measurements are made in projected space.
As a result, the dark matter halos selected and ranked upon those observables display a collective mass measurement biases.  This bias is also predicted in models that account for the observable-mass correlation and the mass measurement scatter. 

Our study differs from previous ones, for example \cite{2020MNRAS.496.4468S}, as we prioritize simplicity and a direct modeling solution over fidelity to the cluster finding process. In our assumption that cluster galaxies are modeled by dark matter subhalos in the massive halo, we preserve the shape and orientation of each halo. Our procedure illuminates  the effect of projection along the line of sight on both the  subhalo counts and measured halo mass.

This analysis provides insight into the selection effect that plagues recent cluster cosmology analyses.
Although the efficacy of the forward-modeling approach in our work still need to be tested with observational data sets and a set of more realistic simulations with light-cone realization and high-resolution mass measurement observable, our analysis demonstrates that the selection effect can be predicted with assumptions about the the correlation between the selection observable and the cluster weak lensing mass measurement deviations. It would be prudent to further understand the physical origins of those correlations, and develop a well-motivated functional form for the correlation. We recommend developing cluster selection observables that are less affected by projected observations on the plane of the sky, thus reducing the level of correlations between cluster selection and weak lensing mass measurement.




\section{Acknowledgements}

We thank the anonymous referee for the very helpful comments and suggestions.
This manuscript has been authored by Fermi Research Alliance, LLC under Contract No. DE-AC02-07CH11359 with the U.S. Department of Energy, Office of Science, Office of High Energy Physics. The IllustrisTNG simulations were undertaken with compute time awarded by the Gauss Centre for Supercomputing (GCS) under GCS Large-Scale Projects GCS-ILLU and GCS-DWAR on the GCS share of the supercomputer Hazel Hen at the High Performance Computing Center Stuttgart (HLRS), as well as on the machines of the Max Planck Computing and Data Facility (MPCDF) in Garching, Germany.

\section{Data Availability}

The data underlying this article were accessed from the Illustris-TNG database. The derived data generated in this research will be shared on reasonable request to the corresponding author.



\bibliographystyle{mnras}
\bibliography{example} 

\clearpage



\bsp	
\label{lastpage}
\end{document}